\documentclass{aastex}

\def\jref#1 #2 #3 #4 {{\par\noindent \hangindent=2em \hangafter=1
      \advance \rightskip by 0em #1, {\it#2}, {\bf#3}, #4.\par}}
\def\rref#1{{\par\noindent \hangindent=2em \hangafter=1
      \advance \rightskip by 0em #1.\par}}
\def\lsim{\,\lower2truept\hbox{${< \atop\hbox{\raise4truept\hbox{$\sim$}}}$}\,}
\def\gsim{\,\lower2truept\hbox{${> \atop\hbox{\raise4truept\hbox{$\sim$}}}$}\,}

\slugcomment{Submitted to ApJ, 14 March 2002.}
\shorttitle{Dynamical effects of the neutrino gravitational clustering
at {\sc Planck} angular scales}
\shortauthors{Popa L.A. et al.}

\begin{document}


\title{Dynamical effects of the neutrino gravitational clustering
at {\sc Planck} angular scales}


\author{L.A. Popa\altaffilmark{1}, C. Burigana and N. Mandolesi}
%\email{aastex-help@aas.org}

%\affil{Institute of Space Sciences, Bucharest-Magurele, R-76900, Romania}

\affil{IASF/CNR, Istituto di Astrofisica Spaziale e Fisica Cosmica,
Sezione di Bologna,\\
Consiglio Nazionale delle Ricerche, Via Gobetti 101, I-40129 Bologna, Italy}
%\email{aastex-help@aas.org}


%\altaffiltext{1}{further address:
%Istituto TeSRE, Consiglio Nazionale delle Ricerche, Via Gobetti 101,
%I-40129 Bologna, Italy}
\altaffiltext{1}{further address:
Institute of Space Sciences, Bucharest-Magurele, R-76900, Romania}


\begin{abstract}

We study the CMB anisotropy
induced by the non-linear perturbations in
the massive neutrino density
associated to the non-linear
gravitational clustering process.
By using N-body simulations, we compute
the imprint left by the  gravitational
clustering  on the CMB anisotropy
power spectrum for all non-linear scales taking into account
the time evolution of all non-linear density perturbations,
for a flat $\Lambda$CHDM model consistent with LSS data
and latest CMB measurements
for different
neutrino fractions $f_{\nu}$ corresponding to a neutrino total
mass in the range allowed by the neutrino oscillation and
double beta decay experiments.

We find that the non-linear time varying  potential
induced by the gravitational clustering process
generates metric perturbations, 
leaving a decrease in the CMB anisotropy power spectrum 
of amplitude $\Delta T/T \approx 10^{-6}$ 
for angular resolutions between $\sim 4$ and 20 arcminutes, depending
on the cluster mass scale and the neutrino fraction $f_{\nu}$.
We find also that the consistency among BOOMERANG, MAXIMA-1 and DASI
CMB angular power spectra and the errors on
most of the cosmological parameters improve when
the non-linear effects induced by the gravitational
clustering are taken into account.
Our results show that,
for a neutrino fraction in agreement with that
indicated by the astroparticle and nuclear physics experiments
and a cosmological accreting mass comparable
with the mass of known clusters,
the angular resolution and sensitivity 
of the CMB anisotropy measurements from the {\sc Planck} satellite
will allow the detection of the dynamical effects of the
gravitational clustering. 

This work has been done
in the framework of the {\sc Planck} LFI activities.
\end{abstract}




\section{Introduction}

The study of the Cosmic Microwave Background (CMB) radiation holds the
key of understanding the seeds of the cosmological structures of our
present universe, enabling the measurements for most of the important
cosmological parameters. The new generation of CMB experiments as
MAP\footnote{http://map.gsfc.nasa.gov} and
{\sc Planck}\footnote{http://astro.estec.esa.nl/Planck}
will achieve enough precision to reveal more cosmological information
on the structure formation process up to arcminute angular scales.\\
Inside the  horizon,  the acoustic, Doppler, gravitational redshift and
photon diffusion effects combine to determine the angular power
spectrum of  CMB primary anisotropies.
The secondary effects generated between the recombination
and the present can also alter the CMB anisotropy, providing more details
on the evolution of the structures and less robust constraints on the
background parameters (Hu, Sugiyama \& Silk 1997).
The secondary effects are basically divided into two categories:
gravitational effects from metric distortions and rescattering effects
from reionization. The gravitational effects include the early
integrated Sachs-Wolfe (ISW)
effect (see e.g. Hu \& Sugiyama 1995), the late ISW effect (Kofman \& Starobinskii
1985), the Rees-Sciama effect (Rees \& Sciama 1968) as well as the
contribution from
gravitational waves and gravitational lensing (Blanchard \& Schneider
1987, Cole \& Efstathiou 1989).
The rescattering effects both erase the CMB primary anisotropies
and generates secondary anisotropies through the Doppler effect (Kaiser 1984).
Secondary anisotropies at very small scales can be
more efficiently generated by higher     order contributions
than by the Doppler effect.
They include the Vishniac effect (Ostriker \& Vishniac 1986, Vishniac 1987)
associated to linear perturbations in the baryon density,
spatial inhomogeneities of the ionization fraction (Aghanim \& Forni 1999, 
Gruzinov \& Hu 1988)
and the kinematic and thermal Sunyaev-Zel'dovich effect
from clusters (Sunyaev \& Zel'dovich 1972).

In this paper we study the CMB secondary anisotropies
induced by the non-linear perturbations in the massive neutrino density
associated to the non-linear gravitational clustering.
The dynamical role of the massive neutrinos in the gravitational clustering
process was extensively considered in the literaure
(Tremaine \& Gunn 1979, Bond, Efstathiou \& Silk 1980,  Bond \& Szalay 1983,
Ma 2000, Primack \& Gross 2000).
The extent to which the massive neutrinos can cluster gravitationally
depends on their mass and the parameters
of the fiducial cosmological model
describing
the present universe.\\
The atmospheric neutrino experiments (Fukuda et al.
1998, Ambrosio et al. 1998) provides strong evidences of neutrino
oscillations implying  a non-zero neutrino mass
with a lower limit in the range $0.04-0.08$ eV.
The latest measurements of the contribution of the double beta decay
to the neutrino mass matrix (Klapdor-Kleingrothaus 2001) placed
an upper limit on the neutrino mass of $m_{\nu} \leq 0.26$eV.
The direct implication of these results is the
non-negligible contribution of the hot dark matter (HDM)
to  the total mass density of the universe and the existence of
three massive neutrino flavors
(i.e. a density parameter $\Omega_{\nu}h^2 \approx\sum_{i=1}^{3}m_i/93$eV,
where $h=H_0/100$ Km s$^{-1}$ Mpc$^{-1}$ is the dimensionless Hubble constant),
that is also required
by the consistency between the CMB anisotropy at the small
scales with the Large Scale Structure (LLS) of the universe  derived by
the galaxy surveys (e.g. Scott \& White 1994; White et al.~1995;
Primack et al.~1995; Gawiser \& Silk 1998).

Evidences has been accumulated that we live
in a low matter density universe (see e.g. Fukugita, Liu \& Sugiyama 1999
and the references therein). Indications like Hubble diagram of Type 1a
supernovae (Riess et al. 1998, Perlmutter et al. 1998) and
the acoustic peak distribution in the CMB anisotropy power spectra
(Hancock et al. 1998, Efstathiou et al. 1999), point to a
universe close to be flat because of a significant cosmological constant
$\Lambda$ possibly due to a large vacuum energy density.
The analysis of the latest CMB anisotropy data
including BOOMERANG (Netterfield et al. 2001),
DASI (Halverson et al. 2001)
MAXIMA-1 (Lee et al. 2001),
and CBI (Padin et al. 2001), alone or complemented with
other cosmological data sets involving galaxy clustering and Lyman Alpha
Forest (Wang, Tegmark \& Zaldarriaga 2001) point also to a
$\Lambda$-dominated low matter density universe with neutrino masses in the
range of $0.04-4.4$ eV. \\
It is therefore worthwhile to investigate the effects of the
gravitational clustering on the CMB angular power spectrum
in $\Lambda$ cosmological models involving a HDM component
($\Lambda$CHDM) in the form of three massive neutrino flavors with
the total mass in eV range.

The linear perturbation theory describes
accurately the growth of density fluctuations from the early Universe
until a redshift $z \sim 100$
(see the CMBFAST code by Seljak \& Zaldarriaga 1996).
The solution involves the integration of
coupled and linearized Boltzmann, Einstein and fluid equations
(Ma \& Bertschinger 1995) that describes the time evolution of the metric
perturbations in the perturbed density field and the time evolution of the
density fields in the perturbed space-time for all the relevant
species (e.g., photons, baryons, cold dark matter, massless
and massive neutrinos). At lower redshifts  the gravitational
clustering becomes a non-linear process
and the solution  relies on numerical simulations. \\
Through numerical simulations, we compute the CMB
anisotropy in the non-linear
stages of the evolution of the
universe when clusters and superclusters of galaxies start to form
producing a non-linear gravitational potential varying with time.
By using a standard particle-mesh method (Efstathiou \& Eastwood 1991,
Hockney \& Eastwood 1981),
we analyze the imprint of the dynamics of the
neutrino gravitational clustering on the CMB anisotropy
power spectrum in a  flat $\Lambda$CHDM model 
(see Tab.~1) with
$\Omega_m$=0.38, $\Omega_{\Lambda}$=0.62,
H$_0$=62~Km~s$^{-1}$~Mpc$^{-1}$ and different
neutrino fractions
$f_{\nu}=\Omega_{\nu}/(\Omega_b+\Omega_{cdm})$=0.06, 0.11, 0.16
corresponding to
$\Omega_{\nu}$=0.022 ($m_\nu$=0.78eV), 0.037 ($m_\nu$=1.35 eV),
0.053 ($m_\nu$=1.89eV).
We assume three massive neutrino flavors,
a primordial power spectrum with the scalar
spectral index n$_{s}$=0.98, an optical depth to the last
scattering $\tau$=0.12 and neglect the contribution
from the tensorial modes (gravitational waves).
This model is consistent with the LSS data
and the CMB anisotropy latest measurements,
allowing in the same time a pattern of neutrino masses
consistent with the results from neutrino oscillation
and double beta decay experiments.\\

\begin{table}[]
\caption[]{Cosmological parameters assumed in this study 
for the considered fiducial flat $\Lambda$CHDM model.}
\begin{flushleft}
\begin{tabular}{ccccccccccc}
$\Omega_{tot}$&$f_{\nu}$&$\Omega_{cdm}$ & $\Omega_b$&$\Omega_{\Lambda}$
&$\Omega_{\nu}$&$m_{\nu}$(eV/flavor)&h&$n_s$&$\tau$\\
\hline
1 &0.06&0.308&0.05&0.62&0.0216&0.26&0.62&0.98&0.12\\
1 &0.11&0.293&0.05&0.62&0.0373&0.45&0.62&0.98&0.12\\
1 &0.16&0.277&0.05&0.62&0.0522&0.63&0.62&0.98&0.12\\\hline
\end{tabular}
\end{flushleft}
\end{table}

In Section 2 we discuss the physical aspects related to the
dynamics of the neutrino gravitational infall
due to non-linear clustering.
Section 3 presents the N-body
simulations used for the computation of the power spectra of the
CMB temperature fluctuations.
In Section 4 we make estimates of the dynamics of the gravitational clustering
imprinted on the CMB angular power spectra by using the latest
anisotropy measurements in the field.
Finally, we draw our conclusions in Section 5.\\
Throughout the paper we use the system of units in which ${\hbar}=c=k_{B}=1$.

\section{The neutrino gravitational infall}

In the expanding universe, neutrinos decouple from the other
species when the ratio of their interaction rate to the expansion rate
falls below unity. For neutrinos with masses in the
eV range the decoupling temperature is T$_{D} \sim$ 1MeV, occurring
at a redsfit z$_{D} \sim $10$^{10}$ (Freese, Kolb \& Turner 1983).
At this time neutrinos behave like relativistic particles with
a  pure Fermi-Dirac
phase-space distribution:
\begin{eqnarray}
f_{\nu}(q,a)=\frac{1} {e^{E_{\nu}/T_{\nu} +1}},
\hspace{0.3cm} E_{\nu}=\sqrt {q^2+a^2 m^2_{\nu}} \, ,
\end{eqnarray}
where: ${\vec q}$ is the neutrino comoving momentum,
${\vec q}=a {\vec p}$, ${\vec p}$ being
the neutrino 3-vector momentum, $E_{\nu}$ is the energy of
neutrino with mass $m_{\nu}$ and $a$ is the cosmic scale factor
($a_0$=1 today).\\
As  neutrinos are collisionless particles, they
can significantly interact with  photons, baryons and cold dark matter
particles only via gravity.
The neutrino phase space density
is constrained by the Tremaine \& Gunn (1979) criterion
that put  limits on the neutrino energy density inside
the gravitationally bounded objects.
Following Tremaine \& Gunn
criterion it is shown that in the  cosmological models
involving a HDM component (the CHDM models)
the compression fraction of neutrinos through a cluster
$f(r)=\rho_{\nu}/\rho_{cdm}$ (where $r$ is the cluster radius)
never exceeds the background
ratio $\Omega_{\nu}/\Omega_{cdm}$ (Kofman et al. 1996).
Because the formation of
galaxies and clusters is a dynamical time process,
the differences introduced in the gravitational
potential due to neutrino gravitational clustering
generate metric perturbations that
affect the evolution  of the density
fluctuations of all the components of the expanding universe.
Fig.~1 presents the  evolution of the
projected mass distributions of cold dark matter plus baryons
and neutrinos obtained from numerical simulations at
few redshift values $z$ (see Section~2 for details of the simulations).
One can see that neutrinos are accreted by the cold dark
matter and baryons, contributing in the dynamic way to the
gravitational clustering process.\\
\begin{figure}

\vspace{-5cm}
\plotone{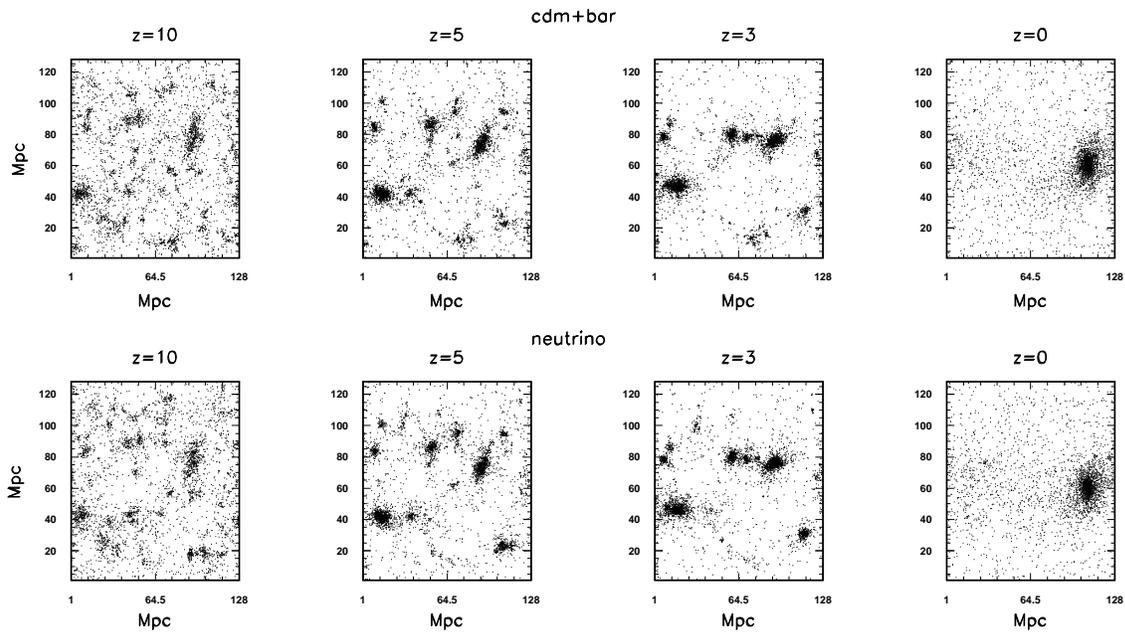}
\caption{The evolution with the redshift of the projected mass distributions of
cold dark matter plus baryons
(upper row) and neutrinos (lower row) obtained from numerical simulation
of $128^3$ cold dark matter plus baryons (the total mass of $8 \times 10^{16} M_{\odot}$)
and $10 \times 128^3$ neutrinos (the total mass of $4.8 \times 10^{18} M_{\odot}$) in a
box of size 128 Mpc, for the $\Lambda$CHDM model with the neutrino
fraction $f_{\nu}=0.16$.  }
\end{figure}
Neutrinos cannot cluster via gravitational instability
on distances below the free-streaming distance $R_{fs}$
(Bond, Efstathiuo \& Silk 1980, Bond \&  Szalay 1983, Ma 2000).
The neutrino free-streaming distance is related to the
causal comoving
horizon distance $\eta(a)$ through (Dodelson, Gates \& Stebbins 1996):
\begin{eqnarray}
R_{fs}(a)=\frac{1}{k_{fs}}=\frac{\eta(a)}{\sqrt{1+(a/a_{nr})^2}}
\,{\rm Mpc},
\hspace{0.35cm}
\eta(a)=\int^a_0 \frac{da}{a^2 H(a)},
\end{eqnarray}
where $a_{nr}$ is the value of the scale factor
when massive neutrinos start to become non-relativistic
($a_{nr}=(1+z_{nr})^{-1} \approx 3k_{B}T_{\nu,0}/ m_{\nu}c^2$)
and $H(a)$ is the Hubble expansion rate:
\begin{eqnarray}
H^2(a)=\frac{8 \pi G}{3}[\Omega_m/a^3+\Omega_r/a^4+
\Omega_{\Lambda}+\Omega_k/a^3].
\end{eqnarray}
In the above equation $G$ is the gravitational constant,
$\Omega_m=\Omega_b+\Omega_{cdm}+\Omega_{\nu}$
is the matter energy density parameter, $\Omega_b$, $\Omega_{cdm}$,
$\Omega_{\nu}$ being the energy density parameters of baryons, cold dark
matter and neutrinos, $\Omega_{r}$ is the
radiation energy density parameter
that includes the contribution from photons and relativistic neutrinos ,
$\Omega_{\Lambda}$ is the vacuum energy density parameter,
$\Omega_k=1-\Omega_m-\Omega_{\Lambda}$ is the energy density parameter
related to the curvature of the universe and $a=(1+z)^{-1}$ is the cosmic
scale factor.\\
\begin{figure}
\plotone{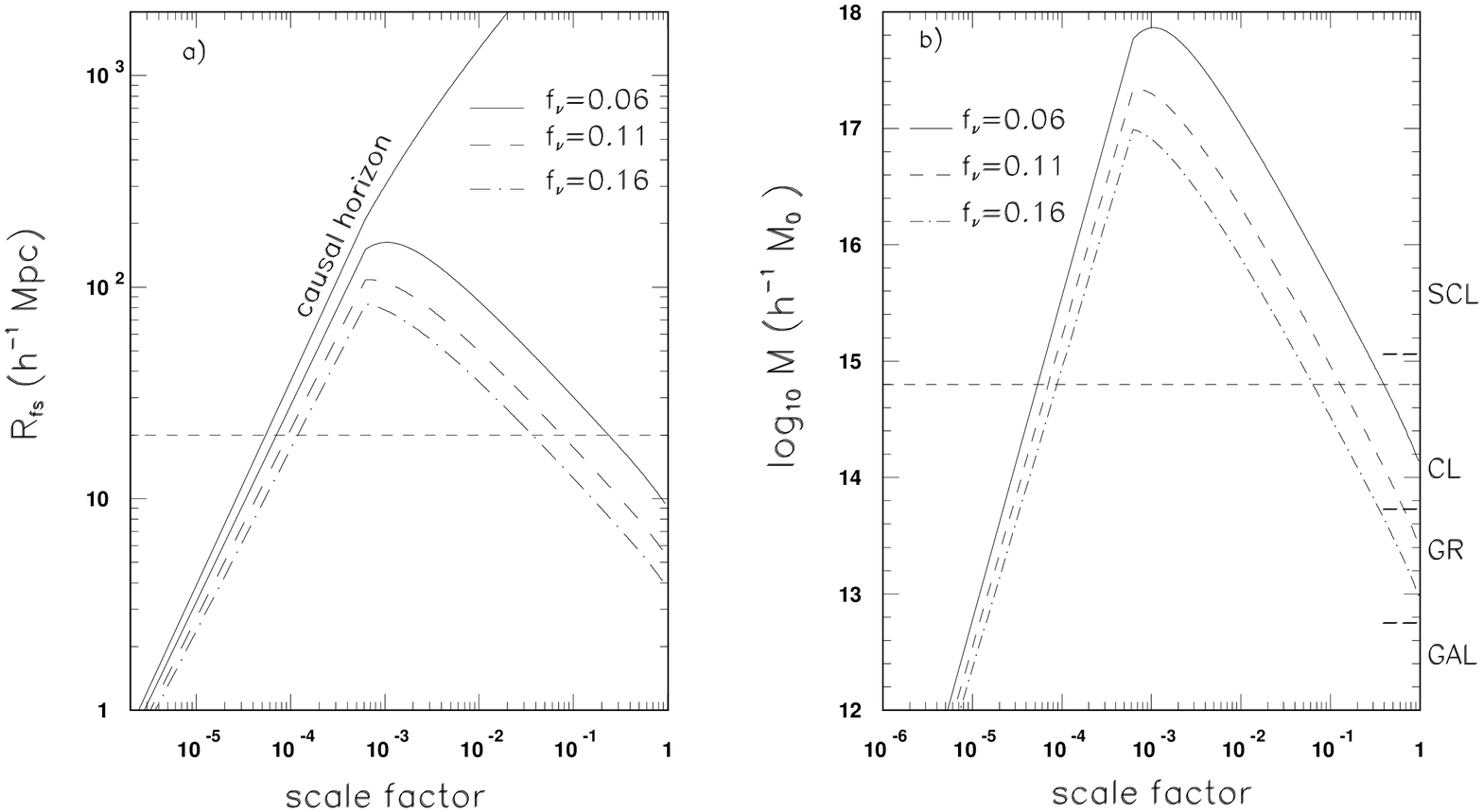}
\caption{Panel a): dependence of the
neutrino free-streaming distance $R_{fs}$ on the scale factor.
We show also a specific scale, $\lambda$=20$h^{-1}$Mpc, constant
in comoving coordinates (horizontal dashed line)
and the evolution
with the scale factor of the causal horizon distance
for our cosmological models.
Panel b): dependence on the scale factor of the
mass $M(R_{fs})$ of the perturbation at the scale $R_{fs}$.
We show also the mass of the perturbation at the scale $\lambda$
(horizontal dashed line) and indicate
the typical mass ranges for galaxies (GAL),
groups (GR), clusters (CL) and superclusters (SCL). }
\end{figure}
$R_{fs}$ defines the minimum linear dimension that a
neutrino perturbation should have in order to survive
the free-streaming.
In the spherical approximation,
the minimum comoving mass of a
perturbation that should contain clusterized neutrinos,
corresponds to (Kolb \& Turner 1990):
\begin{eqnarray}
M(R_{fs}) =\frac{\pi}{6} R_{fs}^3 \rho_m
 \approx 1.5 \times 10^{11}\,(\Omega_m h^2)
(R_{fs}/ {\rm Mpc})^3 h^{-1}M_{\odot}, \nonumber
\end{eqnarray}
where $\Omega_m$ is the matter energy density parameter.\\
We show in Fig.~2
the dependence of the causal horizon distance $\eta(a)$,
 the neutrino free-streaming distance $R_{fs}$,
[panel a)] and of the mass
$M(R_{fs})$ [panel b)] on the cosmic scale factor.
The cosmological model is the $\Lambda$CHDM model
with different neutrino fractions $f_{\nu}$.
One can see that at early times, when neutrinos  are relativistic,
the free-streaming distance is approximately the causal horizon distance.
After neutrinos become non-relativistic ($a_{nr}\sim 10^{-4}$ for
our cosmological models) the free-streaming distance
decreases with time, becoming smaller
than the causal horizon distance.
The time behaviors of $R_{fs}$ and $M(R_{fs})$
show that neutrino can cluster gravitationally on increasingly
smaller scales at latter times.
If the causal horizon $\eta(a)$ is
large enough to encompass the wavelength $\lambda$,
the neutrino gravitational infall perturbs the
growth of the perturbations for this mode,
leaving  imprints in the CMB angular power spectrum.
Perturbations on scales $\lambda < R_{fs}$ ($k > k_{fs}$)
are damped due to the neutrino free-streaming while
the perturbations on scales $\lambda > R_{fs}$ ($k < k_{fs}$)
are affected only by gravity.

\section{CMB anisotropy power spectrum in presence
of neutrino gravitational clustering}

Based on the anisotropy produced by a single cluster
modeled in the spherical approximation,
earlier estimates of the CMB anisotropy
produced by the non-linear density perturbations
(Martinez-Gonzalez \& Sanz 1992)
give only upper limits on degree angular scales.\\
%, ranging
%from $10^{-6}$ to $10^{-5}$.\\
As the  anisotropy produced by the non-linear
density perturbations depends on the
time variations of the
spatial gradients of the gravitational
potential produced by different components
(cold dark matter, baryons, neutrinos),
we calculate the CMB anisotropy in the presence of the
gravitational clustering by using N-body simulation
in large boxes with the side of $128$ Mpc,
that include all non-linear scales used in the computation of
the CMB anysotropy power spectrum from
$\lambda_{min}\approx 12$Mpc ($k_{max}\approx 0.52$Mpc$^{-1}$)
to $\lambda_{max}\approx$ 110 Mpc ($k_{min}\approx 0.06$Mpc$^{-1}$),
taking into account the time evolution of all non-linear
density perturbations influencing the CMB power spectrum
(see also Fig.~1). One should note that $\lambda_{max}$ corresponds to
the comoving horizon size at the matter-radiation equality
for our cosmological models
($\lambda_{eq} \approx 16 \,\Omega_m^{-1}h^{-2}$Mpc).
The non-linear structures are assumed to be formed by
two components: cold dark matter plus baryons and neutrinos in the form
of three massive neutrino flavors,
both components evolving in the gravitational field created
by themself. For the purpose of this work we neglect the
hydrodinamical effects. This approach is justified as
the contributions to the CMB anisotropy of the hot baryonic gas
is proved to be negligible (Quilis, Ibanez \& Saez 1995).

\subsection{Basic equations}

In the Newtonian limit, the neutrino gravitational clustering
can be described as a deviation
from the background by a potential $\Phi$
given by the Poisson equation:
\begin{equation}
{\nabla}^2 \Phi({\vec r},a) =4 \pi G a^2
\rho_m(a)\delta_m({\vec r},a) \, ,
\end{equation}
where ${\vec r}$ is the position 3-vector,
$\rho_m(a)$ is the
matter density
and $\delta_m({\vec r},a)$ is the matter density
fluctuation;
$\delta_m=\delta_b+\delta_c+\delta_{\nu}$, where
$\delta_c$, $\delta_b$ and $\delta_{\nu}$ are the density fluctuations
for cold dark matter particles, baryons and neutrinos. \\
The equations governing the motion of each particle species
(cold dark matter plus baryons and neutrinos)
in the expanding universe
are given by (Kates, Kotok \& Klypin 1991,
Gleb \& Bertshinger 1994):
\begin{eqnarray}
\frac{{d \vec q}}{da}= -a\,H(a) \, {\vec \nabla} \Phi ,
\hspace{0.5cm}  \frac{ d{\vec r} } { da }
={\vec q}\, (a^3 \, H(a))^{-1},
\end{eqnarray}
where ${\vec q}$ is
the comoving momentum and $H(a)$ is the Hubble expansion
rate given by the equation (3).\\
The Newtonian description given by the equations (4)--(5)
applies in the limit of the week gravitational field if, at each time step,
the size of the non-linear structures
is much smaller than the causal horizon
size (the background curvature is negligible).

\subsection{N-body simulations}

The cosmological models involving
massive neutrinos show a characteristic
scale-dependence of the perturbation growth rates
(Ma 1996, Hu \& Eisenstein 1998, Popa, Burigana, Mandolesi 2001).
We evolve the system of baryons plus cold dark matter particles
and neutrinos according to the equations (4) and (5)
for the non-linear scales involved in the computation
of the CMB anisotropy (0.06Mpc$^{-1} \leq k \leq $0.52Mpc$^{-1}$),
starting from the beginning of the non-linear regime
of cold dark matter plus baryons component.\\
The initial positions and velocities of neutrinos
and baryons plus cold dark
matter particles are generated at each
spatial wave number $k$
from the corresponding matter density fluctuations
power spectra at the
present time by using
the Zel'dovich approximation (Zel'dovich 1970).
The matter power spectra was normalized on the basis of the
analysis of the local cluster X-ray
temperature function (Eke, Cole \& Frenk 1996).
We performed simulations with $128^3$ cold dark matter plus
baryon particles and
$10 \times 128^3$ neutrinos.
The neutrinos and the baryons plus cold dark matter particles
was randomly placed
on $128^3$ grids, with  comoving spacing $r_0$ of 0.5~h$^{-1}$Mpc.
The high number of neutrinos and this comoving spacing
ensure a precision high enough for a correct sampling of the neutrino
phase space distribution
(Popa, Burigana, Mandolesi 2001). \\
According to the Zel'dovich approximation, the perturbed comoving
position of each particle ${\vec r}( {\vec r_0}, a)$ and its peculiar
velocity ${\vec v}( {\vec r_0},a)$ are related to the
fluctuations of the density field
$\delta \rho({\vec r_0},a,k)$ through:
\begin{eqnarray}
{\vec r}( {\vec r_0},k,a) &=& {\vec r_0}+D(k,a)
{\vec d}({\vec r_0}) \, , \;\;\;
{\vec v}( {\vec r_0},k,a) = {\dot D}(k,a){\vec d}({\vec r_0}) \, , \\
{\vec \nabla}{\vec d}({\vec r_0})
&=& D^{-1}(k,a) \delta \rho({\vec r_0},k,a) \nonumber \, ,
\end{eqnarray}
where ${\vec r_0}$ is the coordinate corresponding to the
unperturbed comoving position,
${\vec d}({\vec r_0})$ is the displacement field
and $D(k,a)$ is the growth function of perurturbations
corresponding to each cosmological model.\\
At each wave number $k$ used in the computation of the CMB anisotropy
we compute the perturbed particles
comoving positions and  peculiar velocities
at the beginning of the non-linear regime $a_{nl}$, by using
the set of equations (6).
\begin{figure}
%\vspace{-5.cm}
\plotone{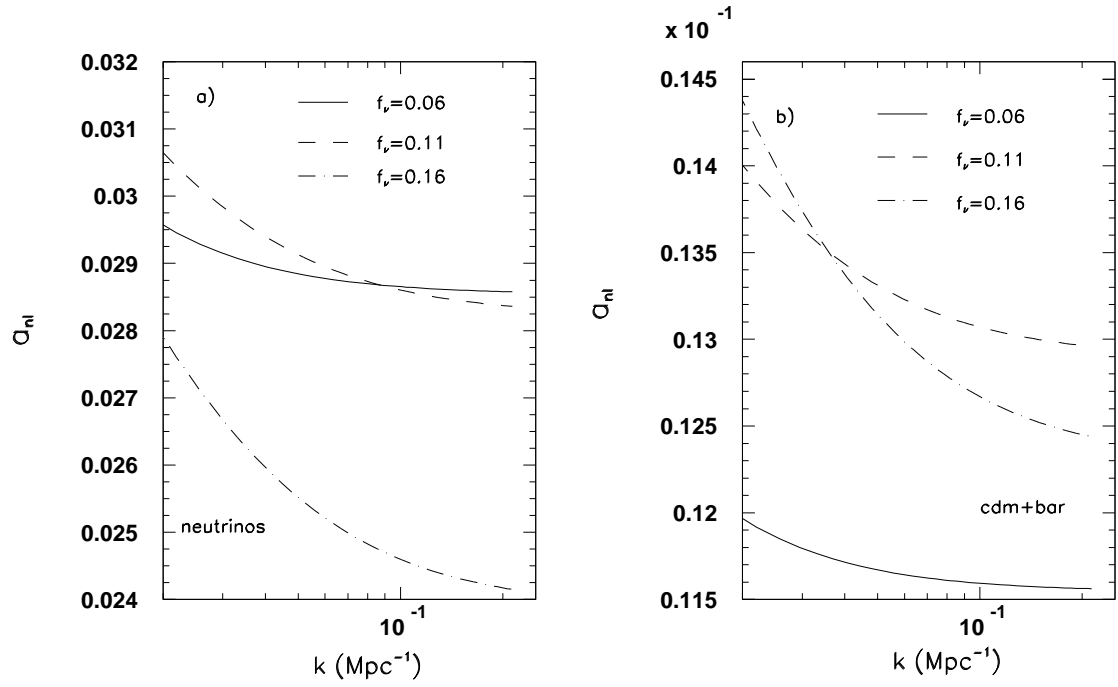}
\caption{The scale dependence of $a_{nl}$
for neutrinos (panel a) and baryons plus cold
dark matter particles (panel b).}
\end{figure}
We assign to each particle a momentum
according to the growth  function,
when the power of each mode is randomly selected
from a Gaussian distribution with the mean accordingly
to the  corresponding power spectrum
(Hoffman \& Ribak 1991, Ganon \& Hoffman 1993, Bertschinger 1995).
In the computation of the set of equation (6) we consider
only the growing modes, the non-linear power spectra up to
$k_{max}=6.28$~h~Mpc$^{-1}$,
and neglect the contribution of the redshift distortions.\\
We present in Fig.~3 the dependence on the spatial wave number $k$
of the scale factor $a_{nl}$ for each component.
One can see from Fig.~3 that
neutrinos (panel~a)  enter in the non-linear regime
later than cold dark matter particles and baryons (panel~b).
Thus, the neutrino halo of the cluster starts to form after
the cold dark matter plus baryon halo is advanced
in the non-linear stage,
causing the accretion of neutrinos from
the background.\\
At each spatial wavenumber $k$ we evolve the
particles positions and velocities according to
the set of equations~(4)-(5).
We start this process  from the scale factor $a^{c}_{nl}$
at which cold dark matter particles plus baryons start
to enter in the non-linear regime.
At each time step, the density on the
mesh is obtained from the particle positions using
the Cloud-in-Cell method and
equation (4) is solved by using 7-point discrete analog
of the Laplacian operator and the FFT technique
(Klypin \& Holtzman 1997). The particle positions and
velocities are then advanced in time with a time step $da$
required by the computation of the CMB anisotropy
power spectra.
\begin{figure}
\plotone{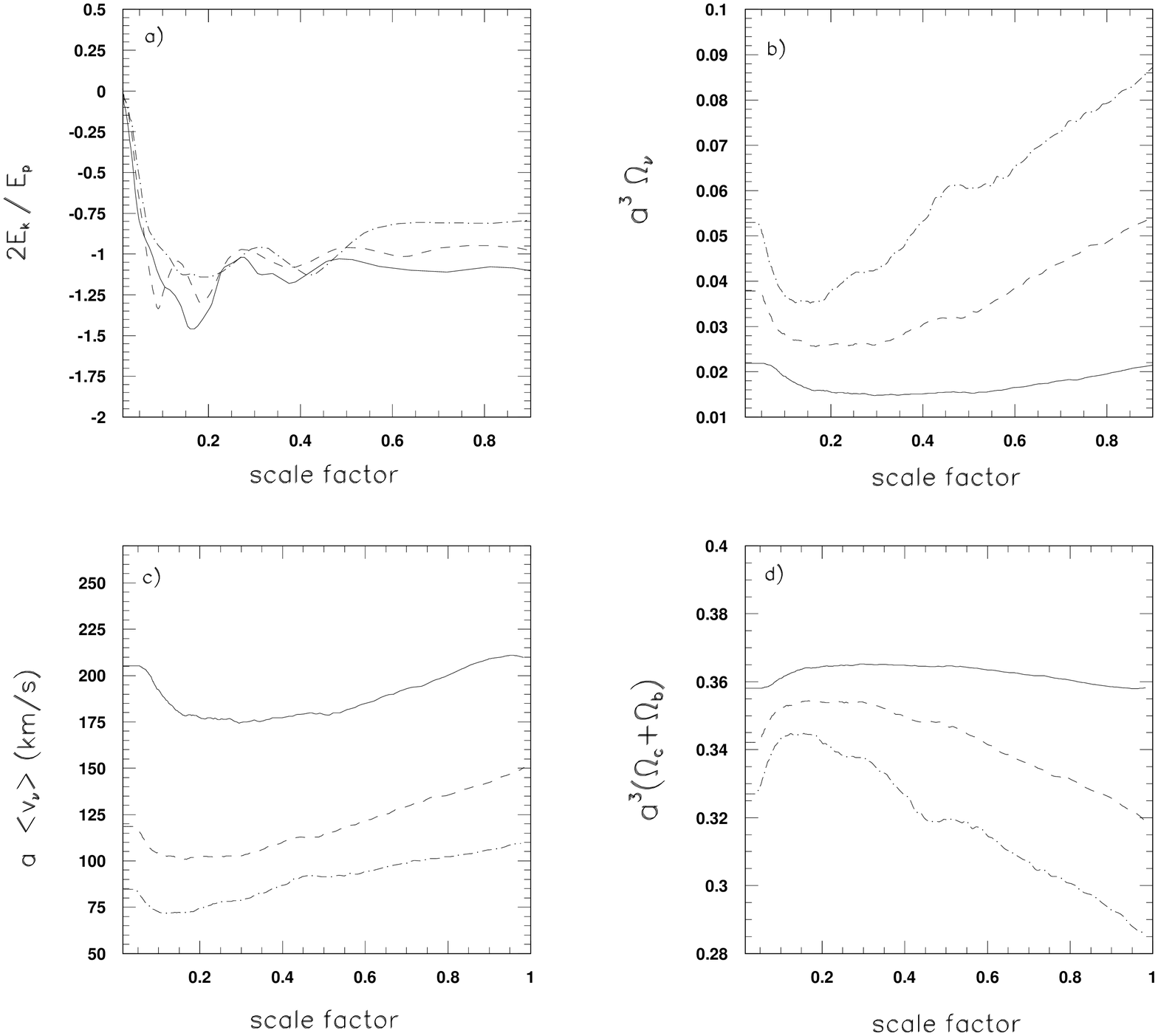}
\caption{The evolution with the scale factor $a$ of:
the ratio of the kinetic energy ($E_{k}$)
and potential energy ($E_{p}$)
for the system of particles considered in the simulation [Panel a)];
the neutrino comoving energy density parameter [Panel b)];
the neutrino averaged comoving velocity [Panel c)];
the comoving energy density parameter of
cold dark matter plus baryons [Panel d)].
All plots are obtained for mode $k=0.06$Mpc$^{-1}$ and different
neutrino fractions: $f_{\nu}$=0.06 (solid lines),
0.11 (dashed lines), 0.16 (dash-dotted lines).}
\end{figure}
The system of particles was evolved until the
scale factor $a_{st}$
when,  according to the virial theorem (Peacock 2001),
it reaches the  equilibrium.
The virial theorem states that
a system of particles evolving in the gravitational field
achieves its equilibrium  state when
the time averaged kinetic energy $E_k$ and the potential energy $E_{p}$
of the system are related through: $E_{p}=-2E_{k}$.
We present in panel~a) of Fig.~4 the evolution
of the ratio $2E_k/E_p$ with the scale factor,
for different neutrino fractions.
In all cases the system of particles achieves the
equilibrium condition at
$a_{st}\approx 0.4$ ($z \approx 1.5$).\\
The neutrino momentum field obtained at each wave number $k$
and scale factor $a$
was sampled in fixed equispaced
points (we use here $N_{q_{max}}$=50) and normalized to
the neutrino total number
($N_{part}=10 \times 128^3$ in the current simulation).
In this way we obtained
the time evolution of the full phase space neutrino distribution
function  by including the perturbation
from the initial phase space distribution
due to the non-linear evolution of the gravitational field.\\
At each time step the neutrino energy density and pressure
are obtained as:
\begin{eqnarray}
\rho_{\nu}(a)=\frac{T^4_{\nu}(a)}{2 \pi^2}\int^{\infty}_{0}
dq \, q^2 E_{\nu}(a)f_{\nu}(a,q),
\hspace{0.4cm}
p_{\nu}(a)=\frac{T^4_{\nu}(a)}{2 \pi^2}\int^{\infty}_{0}
dq \, \frac{q^2}{E_{\nu}(a)}f_{\nu}(a,q),
\end{eqnarray}
where: $T_{\nu}(a)$ is the neutrino temperature
and $E_{\nu}(a)$, $f_{\nu}(a,q)$
are the neutrino  comoving energy and the neutrino
phase space distribution function as given
by the eq.~(1).\\
Panel~b) of Fig.~4
presents the evolution with the scale factor of the
neutrino comoving energy density parameter $a^3\Omega_{\nu}$
obtained for different neutrino fractions.
To better understand the time behavior of the neutrino
energy density parameter, we present in the panel c)
of the same figure the time evolution of the averaged
neutrino comoving velocity $a <v_{\nu}>$.\\
%***********************************************************
When cold dark matter particles plus baryons start
to evolve in the non-linear regime,
neutrinos are still in linear regime, having a pure Fermi-Dirac
phase space distribution. As the system evolves in time,
its potential energy start to increase,
causing perturbations of the neutrino
phase space distribution from the pure Fermi-Dirac
distribution.
Fig.~5 presents the neutrino comoving momentum
distribution functions computed at few different time steps.
For comparison, we plot also a pure Fermi-Dirac distribution.
At early stages, when $a^c_{nl}< a < a^{\nu}_{nl}$
($a^{\nu}_{nl}$ is the value of the scale factor when neutrinos start
to enter in the non-linear regime) neutrinos are not accreted
by the non-linear structures created by the cold dark matter and
baryons as their typical averaged velocity $<v_{\nu}>$ is too large:
\begin{eqnarray}
<v_{\nu}> \approx 160 \,a^{-1}\, \left( \frac{1{\rm eV}}{m_{\nu}}\right)
\, {\rm Km/s}. \nonumber
\end{eqnarray}
As the potential energy of the system increases,
the neutrino averaged comoving velocity, the comoving momentum
and the comoving energy density  are decreased
until the neutrino averaged velocity
drops below the averaged velocity dispersion of cold dark matter
and baryons.
Then, neutrinos start  to fall in the gravitational
potential wells created by the cold dark matter and baryons and
their kinetic energy start to increase. Consequently,
the averaged neutrino comoving velocity, the comoving momentum
and the comoving energy density tends to increase as the system
approaches its equilibrium state.\\
As the consequence of the
matter energy density conservation  each time step,
the comoving energy density of cold dark matter particles plus baryons
shows an opposite time dependence, as presented
in panel d) of Fig.~4.
\begin{figure}
\plotone{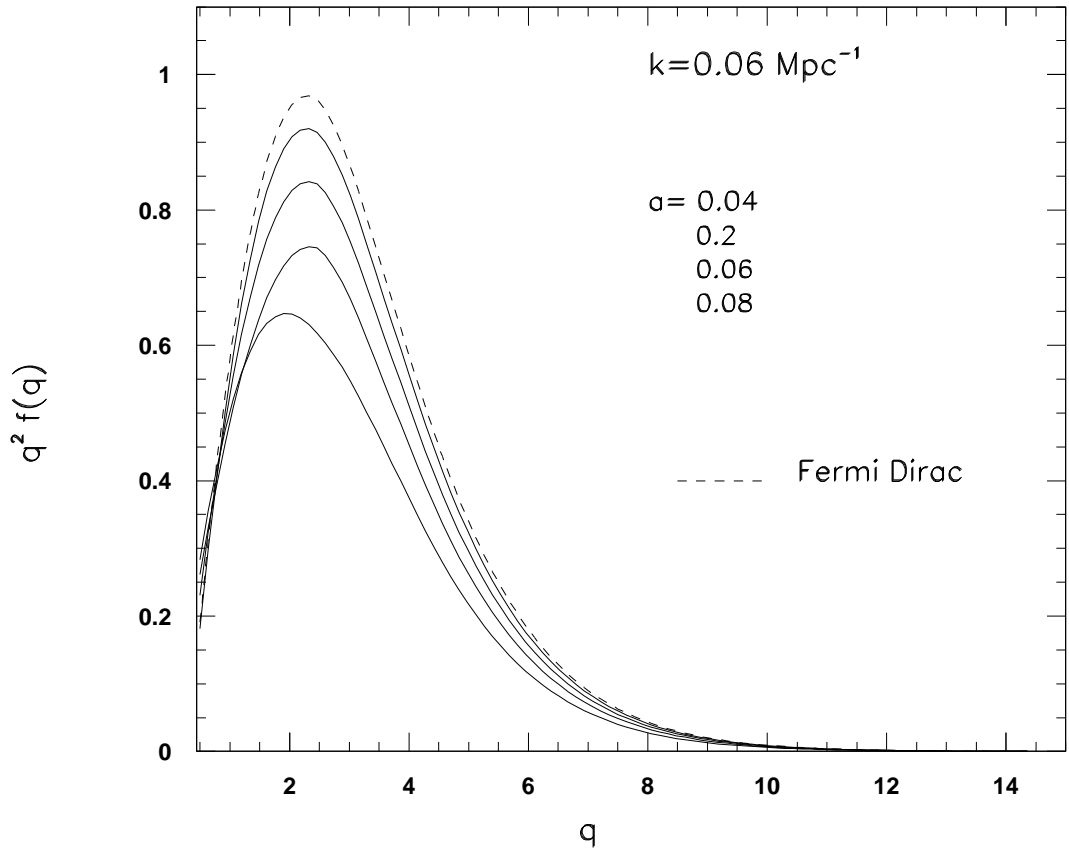}
\caption{The neutrino comoving momentum distribution
function computed at few different time steps. From the top
to the bottom the scale factor is a = 0.04, 0.2, 0.06, 0.08.
The dashed line represents
the  pure Fermi-Dirac distribution function.
All distributions are obtained for the mode k=0.06Mpc$^{-1}$
and for the neutrino fraction $f_{\nu}$=0.06}
\end{figure}

\subsection{Imprints on CMB anisotropy power spectrum}

In this section we compute the
imprints of the dynamics of the neutrino
gravitational clustering on the CMB anisotropy
power spectrum.\\
The neutrino gravitational clustering
can affect both the homogeneous and the inhomogeneous components
of the gravitational field.
The changes in the  homogeneous component
of the gravitational field
are determined by the
changes of the energy density of neutrinos and cold dark matter particles
plus baryons. They affect the Hubble expansion rate,
the sound horizon distance and the neutrino
free-streaming distance [see equations (3) and (4)].
The changes in the  inhomogeneous component
of the gravitational field are determined by the
changes in the energy density for all matter components
and the changes in the neutrino phase space distribution function.
They affect (see equations (21) from Ma \& Bertschinger 1995)
the growth of the energy density perturbations of
cold dark matter, baryons, photons, massive and massless
neutrinos.
\begin{figure}
\plotone{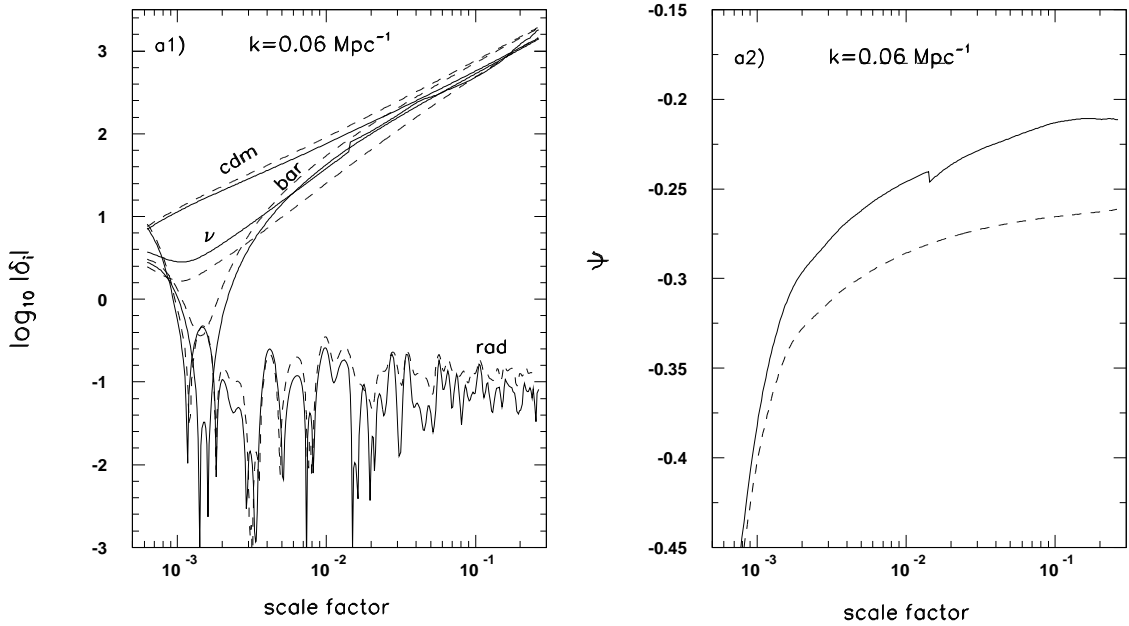}
\caption{Panel a): time evolution of the energy density
perturbations of the different components as computed by including
the gravitational clustering (solid lines) and by neglecting
the gravitational clustering (dashed lines):
cold dark matter (cdm), baryons (bar), massive neutrinos ($\nu$), 
and massless neutrinos plus photons (rad).
Panel b): the same as in panel a), but for
the time evolution of the gravitational field
[k=0.06Mpc$^{-1}$ and $f_{\nu}=0.06$].}
\end{figure}
In the linear regime ($a \leq a^c_{nl}$, for each perturbation
mode  $ k $  we compute the energy density perturbations,
the pressure,
the energy flux and the shear stress for all the components
in the synchronous gauge by using the CMBFAST code.\\
In the non-linear regime ($a \geq a^c_{nl}$),
we compute the neutrino phase
space distribution function, the energy density
of cold dark matter,baryons and neutrinos
at each perturbation mode  $k \geq k_{min}$,
as described in the previous section.
Then, following the same procedure implemented in the CMBFAST code,
we compute in the synchronous gauge the perturbations
of the energy density, the pressure,
the energy flux and the shear stress
for  all the components of our cosmological model.\\
Panel a) of Fig.~6 presents the evolution
with the scale factor
of the energy density perturbations
of different components in the non-linear regime,
for the mode $k=0.06$ Mpc$^{-1}$ (solid lines). For comparison,
we plot also (dashed lines) the energy density perturbations
of the different components
obtained for the same mode $k$ in absence
of the gravitational clustering (linear regime).
The cosmological model
has a neutrino fraction $f_{\nu}=0.06$.
One can see that the neutrino gravitational clustering affects
the growth of the energy density perturbations for
all components.
Panel b) of Fig.~6 presents the evolution
with the scale factor of the scalar potential $\Psi$
of the conformal Newtonian gauge line element, that plays
the role of the gravitational potential in the Newtonian
limit  (Bardeen 1980, Kodama \& Sasaki 1984),
by including (solid line) or not (dashed line)
the effect of neutrino gravitational clustering. 
[For the transformation relation between
the scalar potentials of the synchronous gauge
and conformal Newtonian gauge
see equation (18) of Ma \& Bertschinger (1995)].\\

Fig.~7 presents our computed CMB anisotropy power spectra
in the presence of the neutrino gravitational clustering.
The cosmological model is the $\Lambda$CHDM model
having different neutrino fractions.
We note that the far-IR source clustering contribution 
to the extragalactic foreground fluctuations
may contribute significantly to the 
angular power spectrum at $\ell \gsim 100$ 
(Magliocchetti et al. 2001).
On the other hand, this effect 
may be important also about the first Doppler peak and shows 
a remarkable frequency dependence, 
while the neutrino gravitational clustering
effect decreases the power of the secondary Doppler peaks 
with respect to that of the first Doppler peak 
and does not depend on the frequency.
Therefore, an accurate CMB anisotropy experiment with a wide
frequency coverage, like {\sc Planck}, can in principle 
clearly resolve the combined effect of these two 
contributions.\\
%------------------------------------------------------
\begin{figure}
\plotone{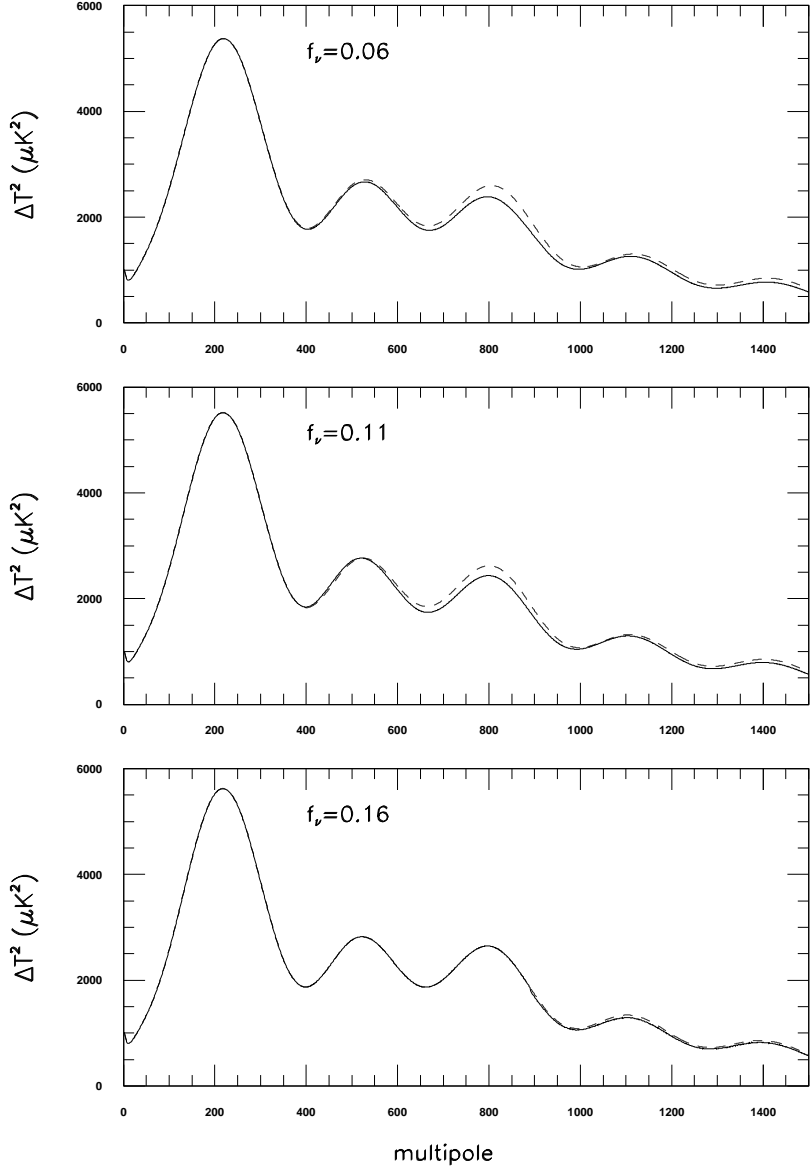}
\caption{The CMB anisotropy power spectra
computed in the presence of the neutrino gravitational
clustering (solid lines).
In each panel we show also the fiducial
$\Lambda$CHDM cosmological model without including
the neutrino gravitational clustering (dashed line),
with: $\Omega_m=0.68$, $\Omega_{\Lambda}=0.62$,
$h=0.62$, $n_s=0.98$, $\tau=0.12$,
three massive neutrino flavors  and different neutrino
fractions ($f_{\nu}$=0.06, 0.011, 0.16, from the top to the bottom).
As usual, we express the CMB anisotropy power spectrum 
in terms of $\Delta T^2 = C_\ell \ell (\ell +1)/2\pi$.
See also the text.}
\end{figure}
%--------------------------
\begin{figure}
\plotone{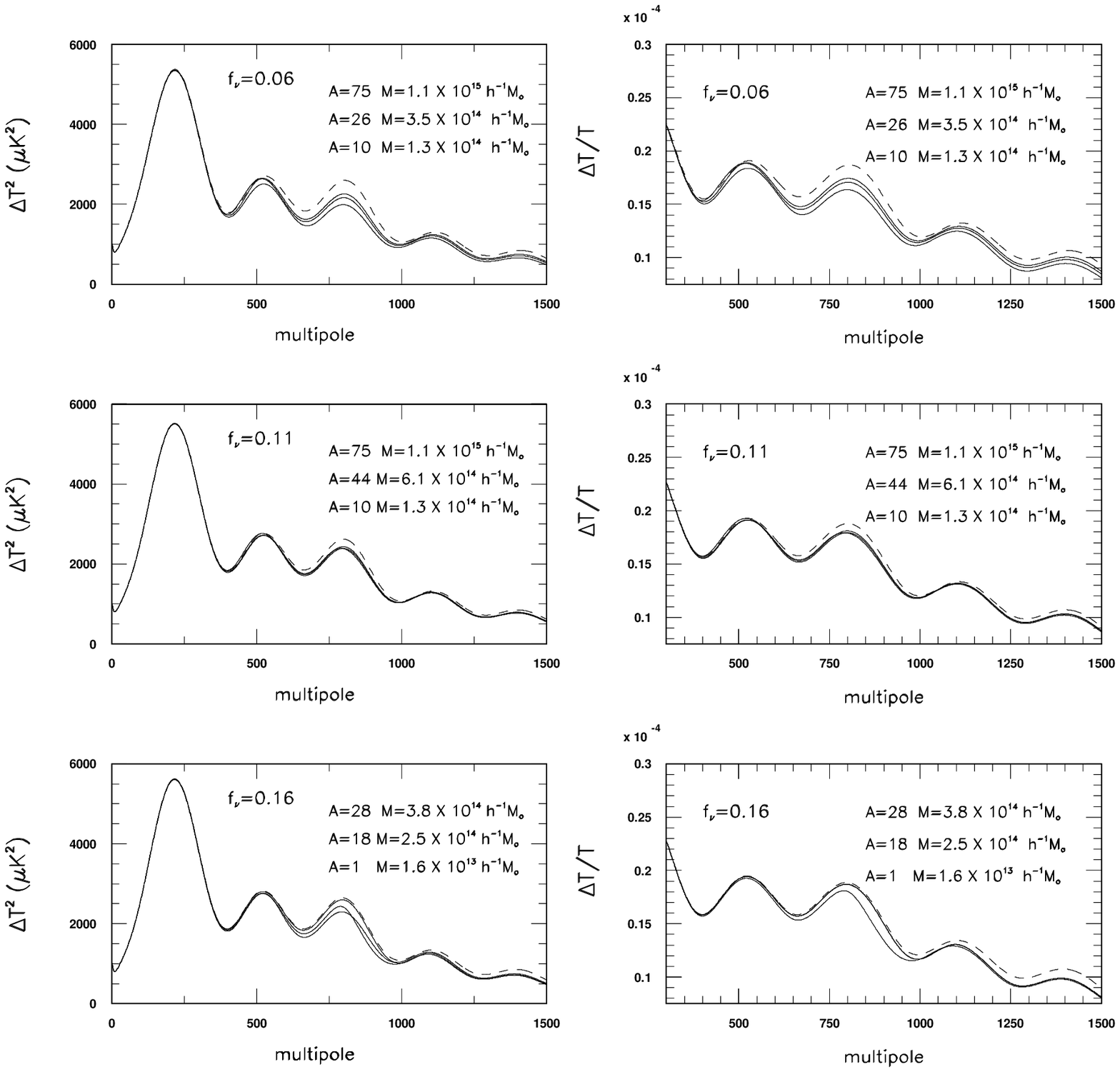}
\caption{Left panels: the  imprint of the neutrino
gravitational clustering on the CMB
anisotropy power spectrum obtained
for the filtering perturbation with the mass $M$.
Right panels: the CMB anisotropy $\Delta T/T$
computed in the presence
of the gravitational clustering
obtained
for the filtering perturbation with the mass $M$.
We report in the panels the richness
of these perturbations.
In each panel, the dashed line corresponds to  the fiducial
$\Lambda$CHDM cosmological model, without including the effect
of neutrino gravitational clustering,  with
$\Omega_m$=0.38, $\Omega_{\Lambda}$=0.62, $h$=0.62,
$\tau$=0.12, $n_s$=0.98, three massive neutrino
flavors  and different neutrino fractions
($f_{\nu}$=0.06, 0.011, 0.16, from the top to the bottom).
See also the text.}
\end{figure}
As we have shown before, the difference in the evolution
of a perturbation mode $k$ depends on how this mode
relates to the neutrino free-streaming wave number $k_{fs}$.
Considering that our simulation at each time step
is a sample of the evolution of
the matter in the non-linear regime,
we study the imprint of the gravitational clustering
on the CMB anisotropy power spectrum
by smoothing the density field
obtained from simulation at each time step
with a filter with the scale $R_{fs}$ corresponding to
the cluster mass value $M(R_{fs})$. For each non-linear mode
$k$ only the perturbations with the mass $M \leq M(R_{fs})$
are  taken into account for the computation of
the CMB anisotropy power spectrum.\\
Fig.~8 presents our computed CMB anisotropy power spectra
obtained when different filtering mass values $M(R_{fs})$
are considered.
It is usual to use the Coma cluster as the mass normalization point
($M_{{\rm Coma}}=1.45 \times 10^{15}$h$^{-1}$M$_{\odot}$);
for the Coma cluster we assume a richness
${\cal A}_{{\rm Coma}}$=106.
According to Kashlinsky (1998),
the relation between the
mass of the perturbation and the richness ${\cal A}$
of the corresponding cluster can be written in the form:
\begin{eqnarray}
M=M_{{\rm Coma}} \frac{{\cal A}}{{\cal A}_{{\rm Coma}}}=
1.45 \times 10^{15}\left( \frac{{\cal A}}{106}
\right) h^{-1}M_{\odot}. \nonumber
\end{eqnarray}
%[We report also in each the panel of Fig.~8 the richness
%of the corresponding filtering perturbation mass].
By comparing the angular power spectra obtained 
including or not the neutrino gravitational
clustering effect, we find a decrease of the CMB angular
power spectrum induced by the neutrino gravitational
clustering of $\Delta T/T \approx 10^{-6}$ 
for angular resolutions between $\sim 4$ and 20 arcminutes, depending
on the cluster mass and neutrino fraction $f_{\nu}$.\\

The characteristic angular scale left by the
the neutrino gravitational clustering on the CMB
anisotropy power spectrum is given by:
\begin{equation}
\theta=\frac{R_{fs}}{\eta_0-\eta(a)},
\end{equation}
where $R_{fs}$ is the scale of the filtering perturbation
with the mass $M(R_{fs})$,
$\eta(a)$ is the particle horizon distance
at the time at which the non-linear perturbation mode
$k$ cross the horizon
and $\eta_0$ is the particle horizon at the present time.
Fig.~9 presents the evolution of the
characteristic scale $\theta$ and of the corresponding
multipole order of the CMB anisotropy power spectrum  with
the mass $M(R_{fs})$.
\begin{figure}
\plotone{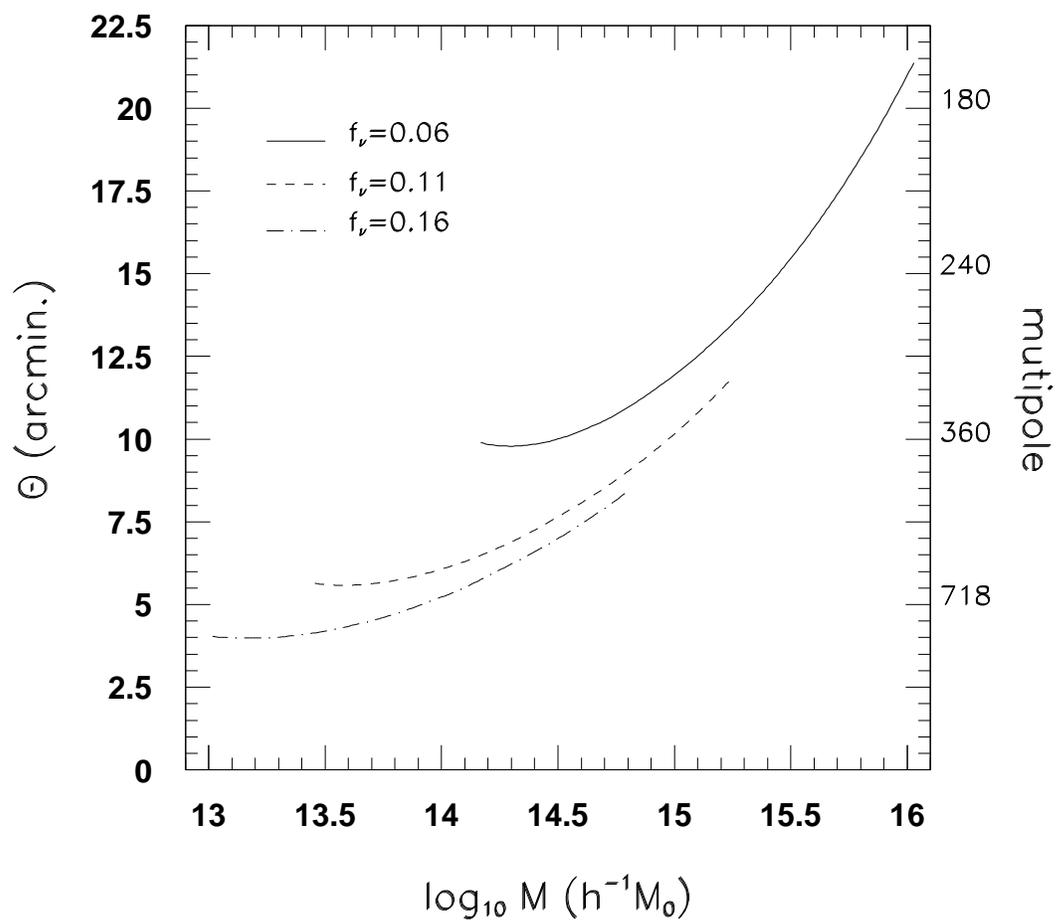}
\caption{Evolution of the characteristic angular scale $\theta$ of
the neutrino gravitational clustering
with $M(R_{fs})$ for different neutrino fractions.
We report also the corresponding multipole orders
$l\sim \theta^{-1}$ of the CMB anisotropy power spectrum.}
\end{figure}

\section{Imprints of neutrino gravitational
clustering at {\sc Planck} angular scales}

The results presented in the previous section show that
not only the high precision satellite experiments as
MAP and {\sc Planck} but also
experiments as BOOMERANG, MAXIMA-1 and
DASI have angular resolutions high enough
to reveal the imprint of the neutrino gravitational
clustering on the CMB angular power spectrum.\\
Of course, high sensitivity anisotropy measurements
are necessary to determine the CMB angular power spectrum
with an accuracy high enough to derive precise
information on the main cosmological parameters
and to be able of disentangle the imprint
of neutrino gravitational clustering.\\
The cosmological parameter estimation from BOOMERANG
data alone or with
various combination of priors was
reported by many authors
(Bond et al. 2000, Lange et al 2001,
Jaffe et al. 2001, Netterfield et al. 2001).
All such limits must be understood
in the context of the specific physical processes that one asks
from the data. Inflation, justified by most of
the experimental measurements in the field, predicts $\Omega_{tot} \simeq 1$
and a scalar spectral index $n_s \simeq 1$.\\
Fig.~10 presents the BOOMERANG, MAXIMA-1 and DASI
angular power spectra together with an inflation best fit model
(Netterfield et al. 2001)
obtained for BOOMERANG
when priors from LSS are taken into account ($\Lambda$CDM model).
In fact, current CMB anisotropy data, complemented with
LSS data, can constrain $\Omega_m$ and $\Omega_b$ from the
morphology of the Doppler peaks.
The cosmological parameters of the considered best fit $\Lambda$CDM
model are:
$\Omega_{tot}=1 \,$,
$\Omega_bh^2=0.021 \,$,
$\Omega_{cdm}h^2=0.13 \,$,
$\Omega_{\Lambda}=0.62 \,$
$\Omega_m=0.37 \,$,
$\Omega_b=0.05 \, $, $h=0.62 \, $ and $\tau=0.12 $.
\begin{figure}
\plotone{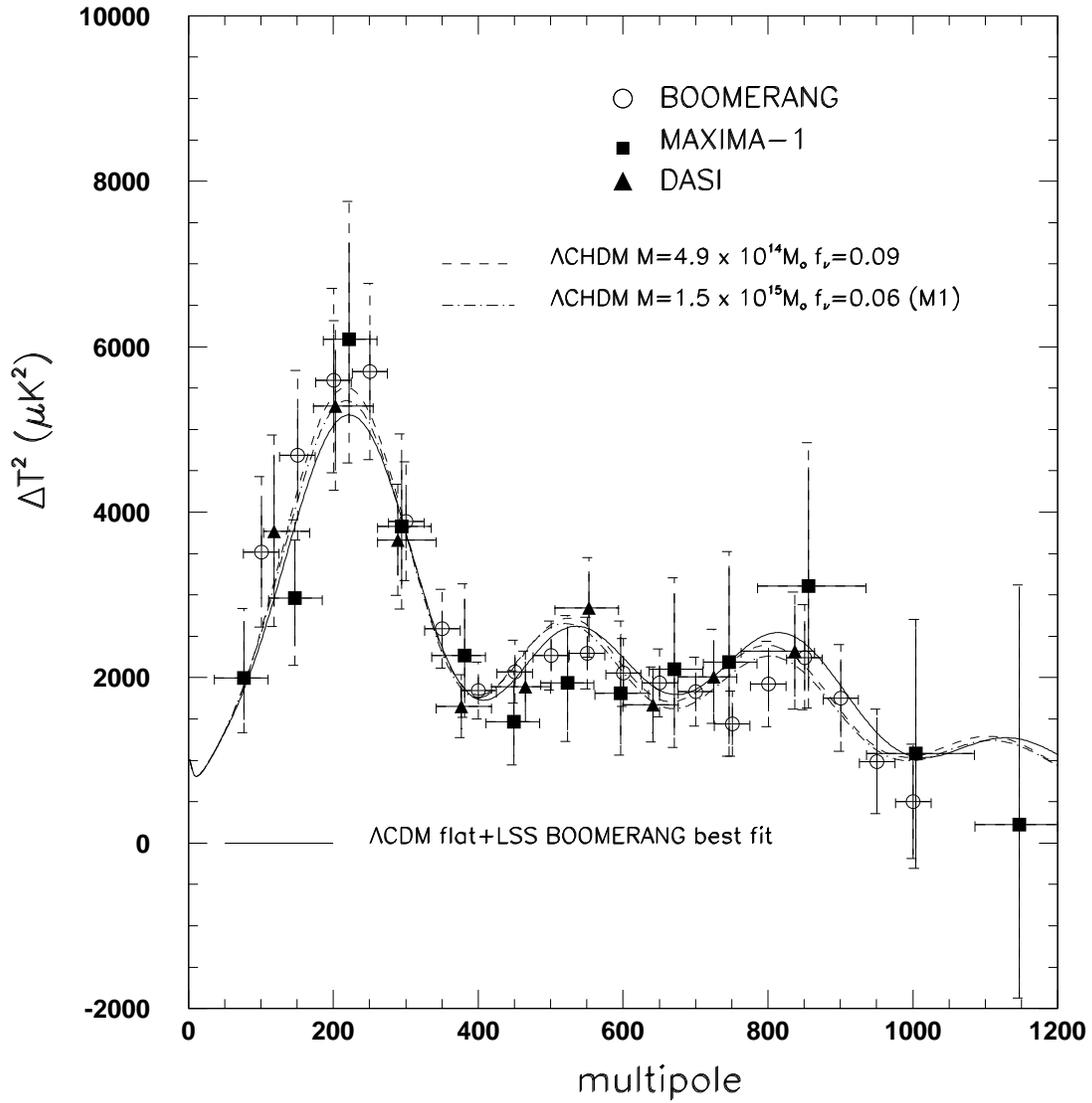}
\caption{The CMB experimental power spectra
form BOOMERANG, MAXIMA-1 and DASI experiments,
the fiducial model
and the best fit models in presence
of the gravitational clustering. The dashed error bars include
also the uncertainty due to systematics and calibration accuracy.
See also the text.}
\end{figure}

We make a simple likelihood analysis of the BOOMERANG,
MAXIMA-1 and DASI measurements in order to see the extent
to which these data sets
can reveal cosmological information
related to the neutrino gravitational clustering
and to test the consistency among these data sets
when the neutrino gravitational clustering is taken
into account. Our computation is justify by the
internal consistency found among these data sets
once calibration, the systematic effects and beam uncertainties
are taken into account (Wang, Tegmark \& Zaldarriaga 2001).\\
For the BOOMERANG experiment, the most relevant systematic
effect is the uncertainty in the beam width
($\delta \sigma_b =
\Delta {\rm FWHM} / \sqrt{8{\rm ln}2} \simeq 1.4'/\sqrt{8{\rm ln}2}$)
while the calibration uncertainty, relevant
for a combined analysis with other experiments,
is of $\simeq 10\%$ (1-$\sigma$ errors; see Netterfield et al. 2001).
We translate
the beam width uncertainty to uncertainty
on recovered CMB angular power spectrum by simply rescaling the
beam window function, $\simeq {\rm exp}[-(\sigma_b \ell)^2]$.
The MAXIMA-1 team directly provides the 1-$\sigma$ errors on CMB angular
power spectrum recovery introduced by the beam width uncertainty and
pointing error, while the quoted 1-$\sigma$ calibration
error is $\simeq 8$\% (Lee et al. 2001).
The DASI angular power spectrum is affected by an uncertainty
of $\simeq 2-4$\% because of the errors in the estimated
aperture efficiency and by a calibration uncertainty of
$\simeq 7$\% (1-$\sigma$ errors; see Halverson et al. 2001
and Leitch et al. 2001).

A detailed analysis of the implications of these experiments
including systematic effects is out of the scope of this work.
We simply include the impact the systematic and calibration errors
by computing upper and lower CMB angular power spectra
by simply rescaling the nominal best fit (i.e. as quoted by the authors
in absence these errors) angular power spectrum data
by including the above multiplicative factor and by respectively
adding or subtracting the statistic errors.\\
%[We neglect in this analysis the two MAXIMA-1 highest multipole data,
%because of their large quoted errors].\\
% Rifatto con tutti: in effetti cambia pochissimo! 

For this analysis we fix the parameters
of our cosmological models to
$\Omega_{tot}=1$, $\Omega_m=0.38$,
$\Omega_{\Lambda}=0.62$, $\Omega_b=0.05$, $h=0.62$, $n_s=0.98$, $\tau=0.12$
and add
different fractions  of  three
massive neutrino flavors  $f_{\nu}$
 in the range 0.001--0.17
(the neutrino total mass in the range 0.001--2~eV).
We define a goodness of the fit as $-2 ln {\cal L}$
in terms of the likelihood function ${\cal L}$
which reduces to the usual gaussian $\chi^2$.
We find that BOOMERANG data are
worst represented when neutrino a fraction is  taken into account.
In this case, the best fit is obtained for a neutrino fraction
$f_{\nu}=0.06 \pm 0.003$ ($m_{\nu}=0.78 \pm 0.039$).
We find an improvement of this fit when the neutrino
gravitational clustering is taken into account,
for a neutrino fraction
$f_{\nu}=0.06 \pm 0.003$ and an accreting mass of
$M=1.5 \times 10^{15} M_{\odot}$ that is closed to the mass of
Coma cluster.\\
Fig.~11 presents the values of the reduced  $\chi^2$
obtained for different data sets and their combinations
for the best fit cosmological model derived from BOOMERANG data.
We also present the reduced $\chi^2$
values obtained for a neutrino fraction $f_{\nu}=0.09$
and an accreting  mass $M=4.9 \times 10^{14} M_{\odot}$,
closed to the mass of the Abell cluster.
The corresponding power spectra are plotted in Fig.~10.
The likelihood values in Fig.~11 are obtained by
taking into account the statistic and systematic errors.
As expected, we find that the values of $\chi^2$ are
reduced (by a factor of about 1.5$-$2, according to
the considered data set) when the calibration errors
of the different experiments are also taken into account.
On the other hand, the basic result remains
the same: the smaller values of $\chi^2$ are
significantly less dependent on the considered set of data
when the neutrino gravitational clustering is included.

We have also computed the errors on the
cosmological parameters by assuming as fiducial model the best fit
derived from BOOMERANG data without clusterization
with a neutrino fraction $f_{\nu}=0.06$
(model M1) and
the same model but including the neutrino gravitational
clusterization with $f_{\nu}=0.06$
and $M=1.5 \times 10^{15} M_{\odot}$ (model M2)
and considering the BOOMERANG data alone (B)
and the combination of the data from BOOMERANG, MAXIMA and DASI (B+M+D).
The result is presented in Tab.~2.
We find that, in general, the errors on the
cosmological parameters are smaller when the neutrino gravitational
clusterization is taken into account.\\
In spite of the accuracy of current CMB anisotropy data,
still quite poor to provide a firm detection
of neutrino gravitational clusterization, and of the
additional uncertainty represented by the relevant systematic
effects, the above results indicate
that including the neutrino gravitational clustering effect
improves the consistency among BOOMERANG, MAXIMA-1 and DASI
CMB angular power spectra, allowing in the same
time a neutrino fraction in agreement with that
indicated by the astroparticle and nuclear physics experiments
and a cosmological accreting mass comparable
with the mass of known clusters.
\begin{figure}
\plotone{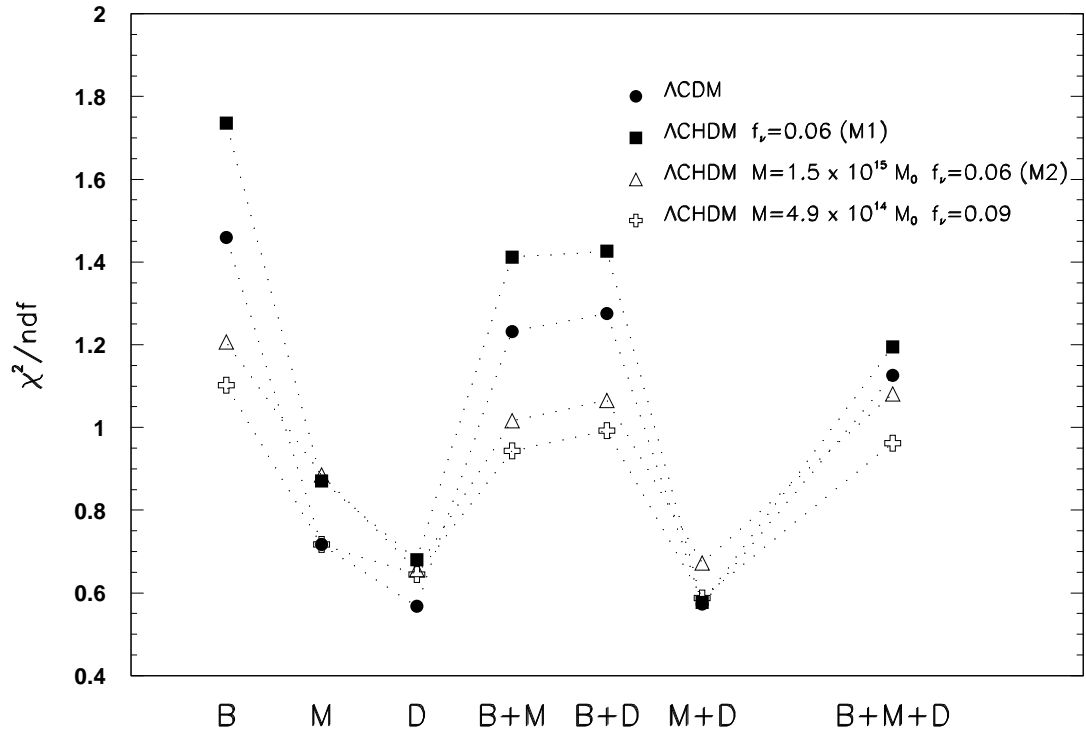}
\caption{The reduced $\chi^2$ obtained for BOOMERANG (B),
MAXIMA-1 (M), DASI (D) and their combinations for the best fit
cosmological models of BOOMERANG data. See also the text.}
\end{figure}
\begin{table}[]
\caption[]{1-$\sigma$ errors on the main cosmological
parameters as obtained by using the Fisher information
matrix taking the covariance matrix from BOOMERANG alone (B)
and the covariance matrix of the combined BOOMERANG, MAXIMA-1 and
DASI experiments (B+M+D). Model M1:
data represented by the $\Lambda$CHDM model with $f_{\nu}=0.06$
by neglecting the effect of clustering.
Model M2: data represented by the $\Lambda$CHDM model
with $f_{\nu}=0.06$ by including the
clustering effect assuming $M=1.5 \times 10^{15}M_{\odot}$.
For the case of B+M+D the covariance matrix was obtained taking into account
the systematic errors and the calibration errors (as indicated in the table).
  }
\begin{flushleft}
\begin{tabular}{cccccccccc}
\hline
Experiment& Model&$n_s$& $\tau$& $\Omega_bh^2$&$\Omega_ch^2$&$\Omega_{\nu}h^2$&$\Omega_m$&$\Omega_{\Lambda}$ &h
\\
B        &M1    &0.083&0.049&0.0031&0.039&0.0045&0.098&0.081&0.087 \\
B        &M2    &0.058&0.021&0.0025&0.026&0.0028&0.081&0.053&0.042\\
B+M+D$^{sy}$&M1 &0.069&0.043&0.0029&0.034&0.0039&0.097&0.047&0.051\\
B+M+D$^{sy}$&M2 &0.051&0.027&0.0026&0.022&0.0018&0.071&0.044&0.036 \\
B+M+D$^{cal}$&M1&0.101&0.071&0.0037&0.057&0.0064&0.141&0.069&0.076\\
B+M+D$^{cal}$&M2&0.089&0.058&0.0031&0.038&0.0031&0.112&0.058&0.062\\
\hline
\end{tabular}
\end{flushleft}
\end{table}

Clearly, new high sensitivity and resolution
space anisotropy experiments will have a much better
capability to detect the neutrino gravitational clustering effect.
In particular, Planck will measure the CMB angular power spectrum
with very high sensitivity up to multipoles $\ell \sim 1000-2000$
with a stringent control of the systematic effects.\\
Fig.~12 presents few confidence regions
of the $f_{\nu} - M$ parameter space that can be potentially detected by
{\sc Planck} surveyor by using the CMB anisotropy measurements
in the presence of the gravitational clustering.
The target model used for this computation
is the $\Lambda$CDM model best fit of the BOOMERANG
data.
We fix the parameters
of our cosmological models to
$\Omega_{tot}=1$, $\Omega_m=0.38$,
$\Omega_{\Lambda}=0.62$, $\Omega_b=0.05$, $h=0.62$, $n_s=0.98$, $\tau=0.12$
and compute the CMB anisotropy power spectra
adding different fractions  of  three massive neutrino flavors  $f_{\nu}$
in the range 0.001--0.1 (the neutrino total
mass in the range 0.001--1.3~eV), for different
 values of the accreting
mass in the range $5 \times 10^{14} - 5 \times 10^{15}$h$^{-1}M_{\odot}$.\\
We consider for this computation only the {\sc Planck} ``cosmological''
channel between 70 and 217~GHz, a sky coverage $f_{sky}=0.8$
and neglect for simplicity the
foreground contamination
(see Popa, Burigana \& Mandolesi 2001 and references therein).\\
At 68\% confidence level we obtain a neutrino
fraction $f_{\nu} \approx 0.011 \pm 0.007$ for an accreting mass
$M \approx (8.2 \pm 3.1) \times 10^{14} h^{-1}M_{\odot}$.
Fig.~11 shows that {\sc Planck} surveyor will have the capability
to measure the imprints of the neutrino gravitational
clustering on the CMB anisotropy power spectrum for a neutrino
mass range in agreement with that
indicated by the astroparticle and nuclear physics experiments
and a cosmological accreting mass comparable
with the mass of the known clusters.
\begin{figure}
\plotone{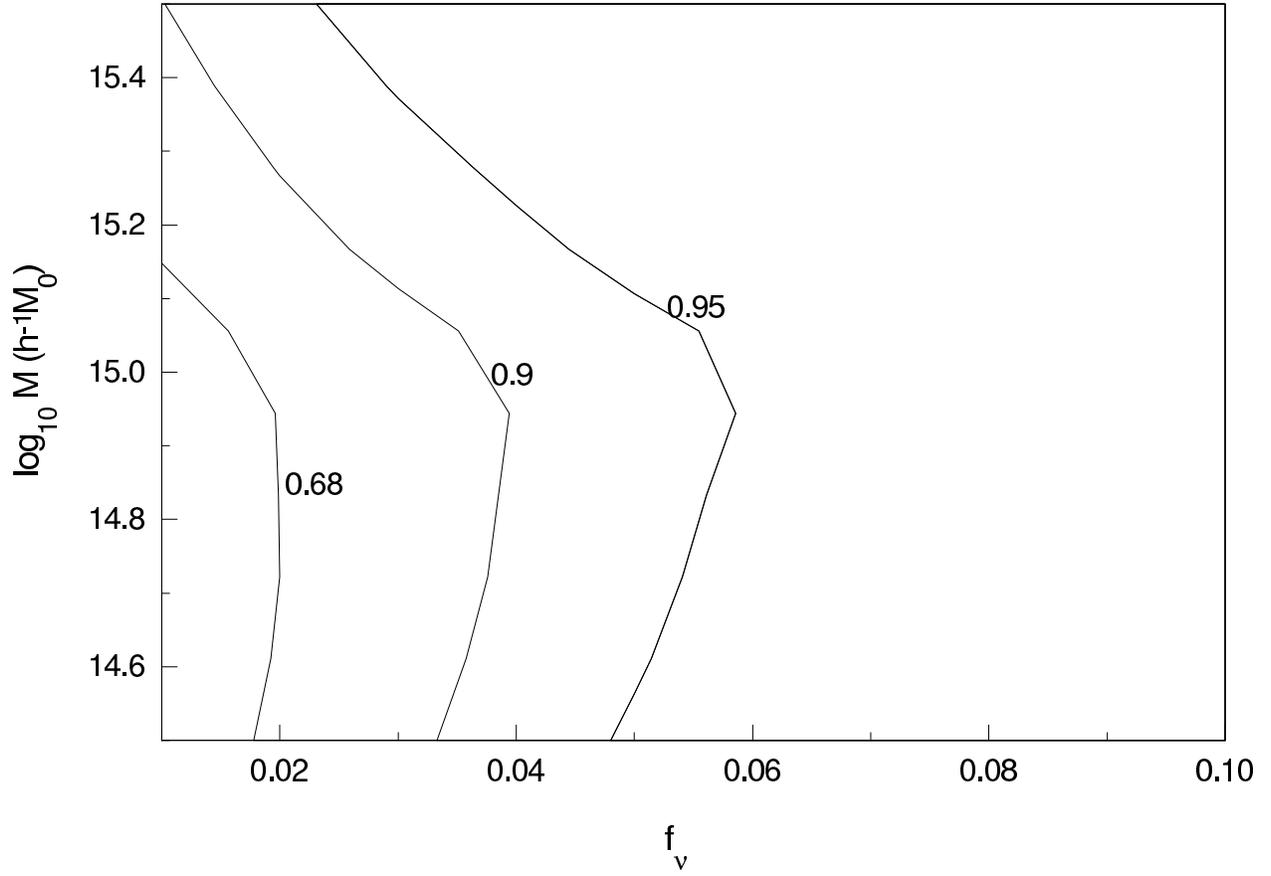}
\caption{Confidence regions of the $f_{\nu}$ - M parameter space
that can be potentially detected by
{\sc Planck} surveyor by using the CMB anisotropy measurements
in the presence of the gravitational clustering. }
\end{figure}

\section{Conclusions}

In this paper we study the CMB anisotropy
induced by the non-linear perturbations in the massive neutrino density
associated to the non-linear gravitational clustering.
Through numerical simulations, we compute the CMB anisotropy
angular power spectrum in the non-linear stages of the evolution of the universe
when clusters and superclusters start to form,
producing a non-linear time varying gravitational potential.
Motivated by the consistency with the LSS data and the latest CMB
anisotropy measurements,
our cosmological model is a flat $\Lambda$CHDM model with different
neutrino fractions $f_{\nu}$ corresponding to a neutrino total
mass in the range allowed by the neutrino oscillation and
double beta decay experiments.\\
We compute the imprint left by the  gravitational
clustering  on the CMB anisotropy
power spectrum for all non-linear scales, taking into account
the time evolution of all non-linear density perturbations
in a large simulation box (with the size of 128 Mpc)
encompassing the comoving horizon size at matter-radiation equality
and a large simulated mass
(the total mass of $\sim 5 \times 10^{18}M_{\odot}$)
that ensures that our simulation is a fair sample of the matter
evolution in the non-linear stages.\\
We found that the non-linear time varying  potential
induced by the gravitational clustering process
generates metric perturbations
that affect the time evolution of the density fluctuations in all the
components of the expanding universe, leaving imprints on the CMB
anisotropy power spectrum at subdegree angular scales.
The magnitute of the induced anisotropy 
and the characteristic angular scale depends on how each non-linear mode
$k$ of the perturbations relates to the neutrino free-streaming wavenumber
$k_{fs}$ at each evolution time step.
By smoothing the density field obtained from simulations
with a filter with the scale corresponding to the
cluster scale,
we find an imprint on the CMB
anisotropy power spectrum of amplitude
$\Delta T/T \approx 10^{-6}$ 
for angular resolutions between $\sim 4$ and 20 arcminutes, depending
on the cluster mass and neutrino fraction $f_{\nu}$.\\
This result suggests that the CMB anisotropy experiments 
with such levels of sensitivities and angular resolutions 
should detect the dynamical effect of
the non-linear gravitational clustering.\\
We analyzed the consistency among BOOMERANG, MAXIMA-1 and DASI
anisotropy measurements when the non-linear effects
induced by the gravitational clustering are taken into account.
Our results suggest 
that including the neutrino gravitational clustering effect
improves the consistency among BOOMERANG, MAXIMA-1 and DASI
CMB angular power spectra and the errors on 
most of the cosmological parameters.
For a neutrino fraction in agreement with that
indicated by the astroparticle and nuclear physics experiments
and a cosmological accreting mass comparable
with the mass of known the clusters,
we find that the {\sc Planck} angular resolution and sensitivity
will allow the detection of the dynamical effects of the
gravitational clustering from the CMB anisotropy measurements.

\acknowledgments

It is a pleasure to thank U.~Seljak and M.~Zaldarriaga
for the use of the CMBFAST code (v3.2) employed
in the computation of the CMB power spectrum and of
the matter transfer function.

\end{document}